# Improved Unet model for brain tumor image segmentation based on ASPP-coordinate attention mechanism

Zixuan Wang[1*], Yanlin Chen[2], Feiyang Wang[3], Qiaozhi Bao[4]

[1]College of Engineering, Carnegie Mellon University, California, Mountain View, 94035, USA.
[2]Tandon School of Engineering, New York University, New York, Brooklyn, 11201, USA.
[3]Information Networking Instutite, Carnegie Mellon University, California, Sunnyvale, 94085, USA.
[4]Department of Statistics, North Carolina State University, North Carolina, Raleigh, 27695, USA.
[*]Corresponding author: e-mail: zixuanwa@alumni.cmu.edu

*Abstract*—In this paper, we propose an improved Unet model for brain tumor image segmentation, which combines coordinate attention mechanism and ASPP module to improve the segmentation effect. After the data set is divided, we do the necessary preprocessing to the image and use the improved model to experiment. First, we trained and validated the traditional Unet model. By analyzing the loss curve of the training set and the validation set, we can see that the loss value continues to decline at the first epoch and becomes stable at the eighth epoch. This process shows that the model constantly optimizes its parameters to improve performance. At the same time, the change in the miou (mean Intersection over Union) index shows that the miou value exceeded 0.6 at the 15th epoch, remained above 0.6 thereafter, and reached above 0.7 at the 46th epoch. These results indicate that the basic Unet model is effective in brain tumor image segmentation. Next, we introduce an improved Unet algorithm based on coordinate attention mechanism and ASPP module for experiments. By observing the loss change curves of the training set and the verification set, it is found that the loss value reaches the lowest point at the sixth epoch and then remains relatively stable. At the same time, the miou indicator has stabilized above 0.7 since the 20th epoch and has reached a maximum of 0.76. These results show that the new mechanism introduced significantly improves the segmentation ability of the model. Finally, we apply the trained traditional Unet model and the improved Unet model based on the coordinate attention mechanism and ASPP module to the test set for brain tumor image segmentation prediction. Compared to the traditional Unet, the enhanced model offers superior segmentation and edge accuracy, providing a more reliable method for medical image analysis with the coordinate attention mechanism and ASPP module.

## I. Introduction

Image segmentation of brain tumor is an important research field in medical image analysis, aiming to accurately extract brain tumor regions from medical images. With the development of modern medical imaging technology, such as magnetic resonance imaging (MRI) and computed tomography (CT), doctors can obtain high-resolution brain images, which provide an important basis for tumor diagnosis, treatment plan and prognosis assessment [1]. However, manual segmentation of brain tumors is not only time-consuming and labor-intensive, but also susceptible to subjective factors of the operator. Therefore, the development of automatic image segmentation methods has important clinical significance.

Brain tumor is a serious threat to human health, its types and forms are different, often in young adults. According to the statistics of the World Health Organization (WHO), the incidence of malignant brain tumors is increasing year by year, so early diagnosis and accurate evaluation are crucial [2]. Second, due to the significant differences between different patients, traditional algorithms are not effective in dealing with tumors with complex shapes and blurred boundaries, which has prompted researchers to seek more advanced methods to improve segmentation accuracy. In addition, with the development of deep learning technology, it provides new ideas for medical image analysis, making automation and intelligence possible.

Deep learning algorithm is widely used in brain tumor image segmentation, and its advantages are mainly reflected in the following aspects. First, convolutional neural network (CNN), as an important part of deep learning, can effectively capture complex patterns and details in images through multi-level feature extraction. When processing high-dimensional medical images, compared with traditional machine learning methods, CNN can better adapt to the characteristics of the data, thus improving the segmentation accuracy. Some network architectures that have emerged in recent years, such as U-Net[3] and Mask R-CNN[4], have been optimized specifically for medical image processing. These models can not only achieve efficient feature extraction, but also have good contextual information capture ability, so as to identify and segment tumor regions more accurately. For example, U-Net enables the network to fuse information at different scales through symmetrical structural design, thus improving the detection capability of small targets (such as small tumors). The application of deep learning algorithm in brain tumor image segmentation not only improves the accuracy of segmentation results, but also provides strong support for clinical decision-making. With automated segmentation tools, doctors can obtain patient lesion information faster, enabling personalized treatment plans to improve patient survival and quality of life.

To sum up, the research background of brain tumor image segmentation is rich and significant, and deep learning algorithms have brought revolutionary progress to this field. In this paper, Unet is improved based on coordinate attention mechanism and ASPP module to improve the segmentation effect of the model in brain tumor images.

## II. DATA FROM DATA ANALYSIS

The data set used in this paper is selected from the open source data set, which contains 1000 brain tumor images and the corresponding 1000 tumor image segmentation results. We divided the data into the training set and the verification set according to the ratio of 7:3, and selected some data sets for display, as shown in Figure 1.

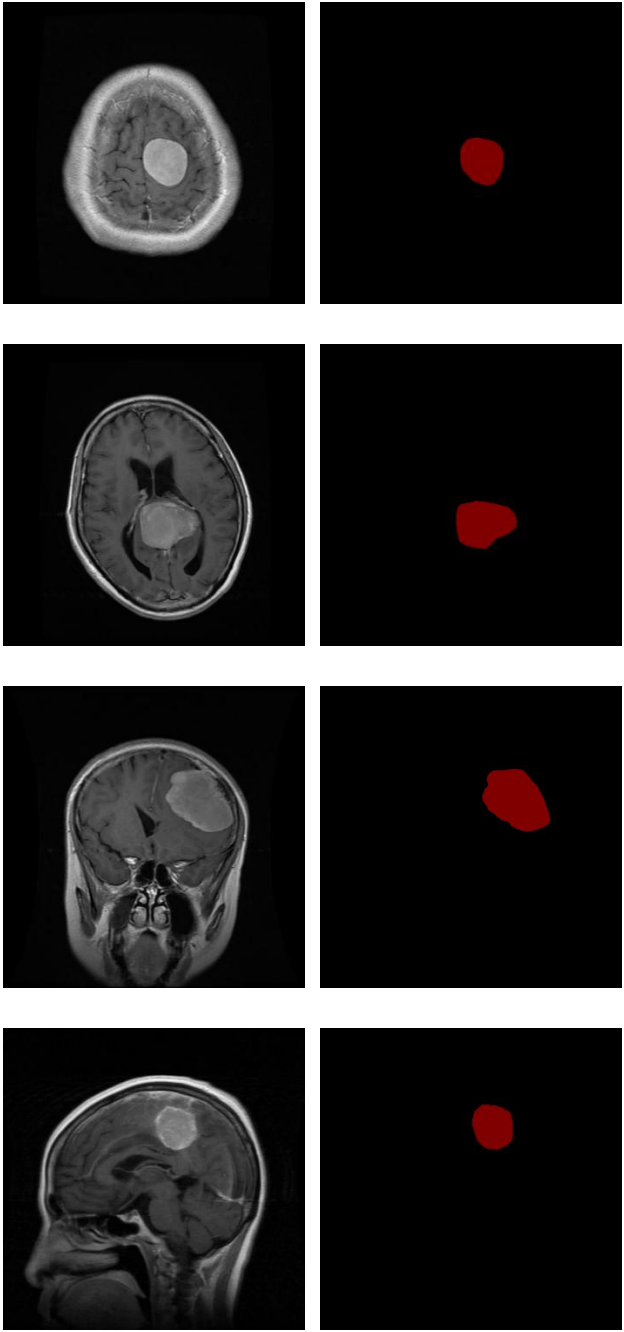

Figure 1. Some of data sets.

## III. IMAGE PREPROCESSING

Before the experiment, we first preprocess the image, including conversion to gray image and histogram equalization.

### A. Convert a color image to a grayscale image

Grayscale image is an image with only brightness information and no color information. The process of converting a color image to a grayscale image uses the weighted average method. In the RGB (Red, Green, Blue) color model, each color channel contributes differently to the human eye's perception of brightness. In general, green contributes the most to brightness, followed by red and then blue.

### B. Image enhancement

Image enhancement aims to improve image quality and make certain features stand out more. Common enhancement methods include histogram equalization, contrast stretching and filtering. This paper uses histogram equalization to enhance the image.

The histogram equalization process is accomplished in the following four steps:

1. Calculate the histogram: calculate the frequency of each gray level.

2. Calculate the cumulative distribution function (CDF) : calculate the cumulative frequency corresponding to each gray level according to the histogram.

3. Normalized CDF: The CDF is normalized to the range [0, L-1] (L is the number of possible grays, usually 256).

4. Map new values: Replace the original grayscale values with normalized CDF to generate new enhanced images.

## IV. METHOD

### A. U-Net

The core structure of U-Net can be viewed as a symmetric encoder-decoder network. It mainly consists of two parts: the encoding path (downsampling) and the decoding path (upsampling). In the coding path, the input image passes through a series of convolution layers and pooling layers for feature extraction and spatial information compression. This process allows the network to capture high-level features in the image while reducing the spatial dimension of the feature map. The model structure of U-net is shown in Figure 2.

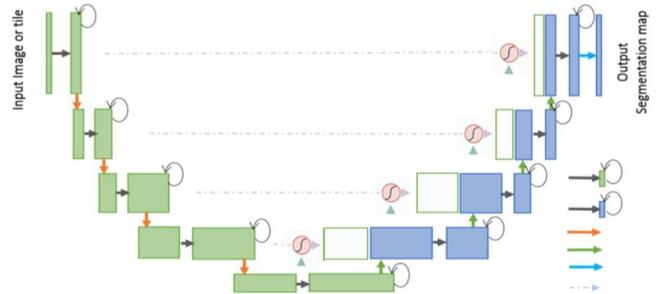

Figure 2. The model structure of U-net .

In the decoding path, U-Net adopts the up-sampling operation to gradually recover the spatial dimension of the feature map by transposing convolution or up-sampling layer. In this process, U-Net introduces skip connections to concatenate the feature map of the corresponding layer in the encoding path with the feature map in the decoding path. This design allows the network to retain more detailed information, which improves the accuracy of the segmentation results.

Jump connectivity is one of the innovations of U-Net. By combining low-level features with high-level features, U-Net is able to effectively fuse information at different scales. This is especially important for edge detection and small target segmentation, as these details are often lost during downsampling. In addition, the jump connection can also alleviate the problem of gradient disappearance, make the model more stable during training, and improve the convergence speed.

During training, U-Net typically uses the cross entropy loss function or Dice coefficient loss function to evaluate the difference between the model output and the real label. Dice coefficient is especially suitable for unbalanced data sets because it can better reflect the segmentation effect of small target regions. In order to improve the generalization ability of the model, data enhancement techniques such as rotation, flipping, scaling, etc. can be used to increase the diversity of training samples.

### B. Coordinate attention mechanism

Coordinate attention mechanism is a technique used to improve the performance of deep learning models when processing images and other spatial data. It mainly introduces spatial location information to enhance the model's ability to focus on specific regions, so as to improve feature extraction and representation. The structure of the coordinate attention mechanism is shown in Figure 3. The work flow of the coordinate attention mechanism is as follows:

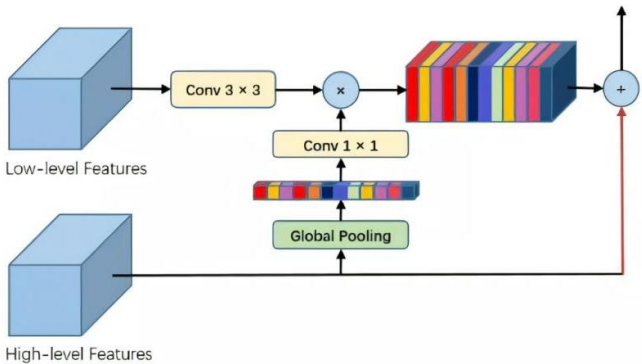

Figure 3. The structure of the coordinate attention mechanism.

1. Spatial information coding: The coordinate attention mechanism encodes the spatial position information of the input feature map to generate position weights that match the size of the feature map. These weights can indicate which locations are more important to the model when processing features.

2. Generate coordinate attention maps: This mechanism usually calculates attention maps in two directions (horizontal and vertical). Specifically, it averages the feature map of each channel globally separately to obtain the corresponding horizontal and vertical coordinate information. After a series of nonlinear transformations of these information, the final coordinate attention diagram is generated.

3. Weighted feature fusion: Finally, by multiplying the original feature map with the generated coordinate attention map element by element, weighted different regions in the feature map are realized. This process allows the model to focus more on important areas while suppressing irrelevant information, thus improving the performance of downstream tasks such as object detection, semantic segmentation, etc.

Coordinate attention mechanism is one of the important research directions in modern computer vision field, which makes deep learning model more efficient and accurate in processing data with spatial structure by introducing the ability of location perception.

### C. ASPP module

ASPP module is a deep learning structure for image segmentation and semantic understanding, which is mainly used to process feature information at different scales. By introducing void convolution, it increases the receptive field while maintaining the resolution of the feature map, thus capturing multi-scale information effectively.

The core idea of the ASPP module is to use multiple cavity convolution layers in parallel, which have different cavity rates. Specifically, ASPP typically consists of several parallel void-convolution layers, each using a different void-rate to extract information from a different spatial range. In addition, ASPP can include a global averaging pooling layer to capture information for the entire image. These feature maps are spliced together and then fused through a 1x1 convolution layer to produce a richer feature representation.

This structure is well suited for semantic segmentation tasks because it is able to efficiently process the various sizes that objects may appear in the image while maintaining sensitivity to detailed information. Therefore, ASPP modules are widely used in many modern deep learning models, such as DeepLab series networks, and have made important contributions to improving segmentation accuracy.

## V. RESULT

After dividing the data set, the image is preprocessed, and then the model in this paper is introduced for experiment. Set the learning rate to 0.0002, batch size to 32, and epoch to 50. In terms of hardware setup, this experiment is based on NVIDIA's RTX 3090 to speed up the training process. In terms of model evaluation parameters, this paper uses loss and miou to evaluate the segmentation effect of the model.

In the field of deep learning and computer vision, especially in image segmentation tasks, Loss and miou are two very important evaluation metrics. The loss function is used to measure the difference between the predicted value of the model and the true value. It is the goal of minimization during optimization. Miou is an important index used to evaluate the performance of image segmentation models. The higher the Miou, the better the performance of the model in the image segmentation task. By combining the loss function and mIoU, the model training and performance evaluation can be effectively guided, thus improving the accuracy and robustness of image segmentation tasks.

First, Unet was used for training and verification, and the loss value change curve of training set and verification set was output, as shown in Figure 4. According to the loss change curves of the training set and the verification set, the loss value decreases from epoch 1 to epoch 8, and then gradually converges regionally.

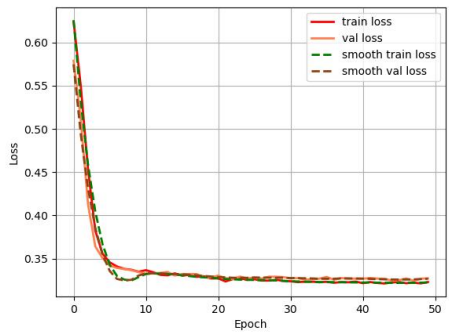

Figure 4. The loss value change curve of training set.

The change of miou during the output training process is shown in Figure 5. According to the miou change curve, the value of miou increased to more than 0.6 at the 15th epoch, and then remained above 0.6 all the time, and miou reached more than 0.7 at the 46th epoch model.

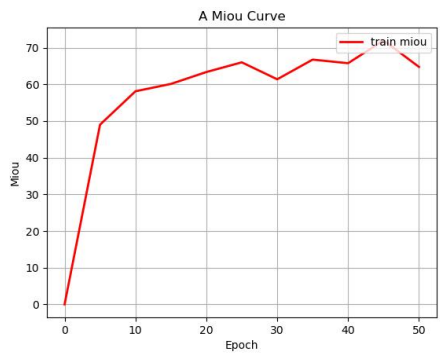

Figure 5. The change of miou during the output training process.

After using unet to train and verify the model, the coordinate attention mechanism and ASPP module in this paper are imported to conduct experiments on the Unet algorithm, and the loss value change curves of the training set and the verification set are output, as shown in Figure 6. According to the loss change curves of the training set and the verification set, the loss value reaches its lowest point at the 6th epoch and then becomes stable.

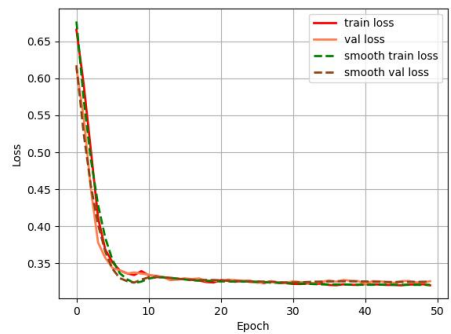

Figure 6. the training set and the verification set.

The changes of miou during the training process are shown in Figure 7. According to the change curve of miou, it can be seen that miou stabilizes above 0.7 and reaches a maximum of 0.76 after the 20th epoch.

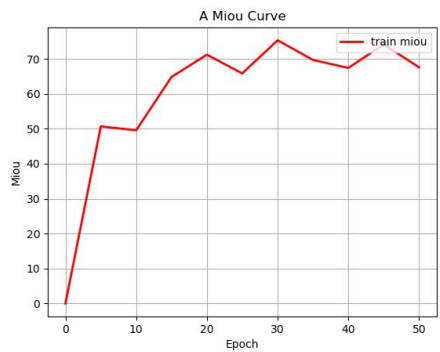

Figure 7. The changes of miou.

The trained Unet model and the coordinate attention mechanism and ASPP modules were respectively used to segment and predict the brain tumor images of the test set by the Unet algorithm model, and the results were shown in Figure 8. The first is the original image, the second is the segmentation result of Unet model, and the third is the segmentation result of Unet algorithm model based on coordinate attention mechanism and ASPP module. Compared with the segmentation results of the two models, we can see that the Unet algorithm model based on coordinate attention mechanism and ASPP module proposed in this paper has better segmentation effect on brain tumor images. Especially in the effect of edge segmentation, the models proposed in this paper show better segmentation ability.

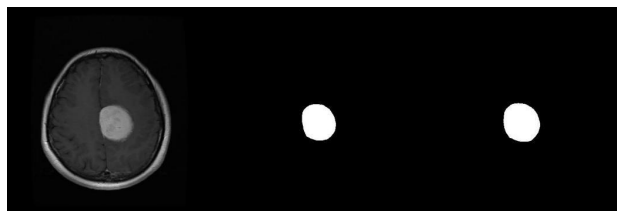

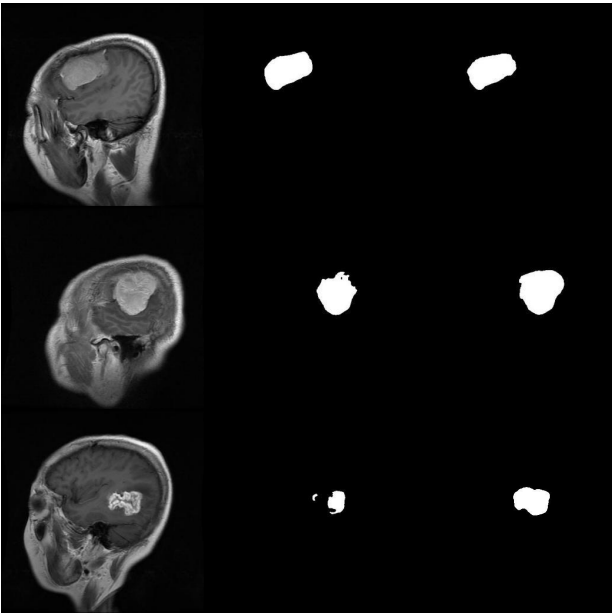

Figure 8. Test set segmentation results.

## VI. CONCLUSION

In this paper, an improved Unet model based on coordinate attention mechanism and ASPP module is proposed for brain tumor image segmentation. In the research process, we first divided the data set reasonably, and carried out the necessary preprocessing of the image, which laid the foundation for the subsequent experiment.

In the preliminary experiment, we use the traditional Unet model for training and verification. According to the loss curve of training set and verification set, the loss value of the model decreases gradually at the first epoch and tends to converge after the eighth epoch. This process shows that the traditional Unet model can effectively learn features in brain tumor image segmentation tasks. At the same time, the change of miou index also reflects the improvement of model performance. At the 15th epoch, the miou value successfully exceeded 0.6 and remained above this level in subsequent training. In particular, at the 46th epoch, miou reaches more than 0.7, which indicates that the traditional Unet model has a certain segmentation ability, but there is still room for further improvement.

To further optimize the model performance, we introduce an improved Unet algorithm based on coordinate attention mechanism and ASPP module. Through comparative experiments, we observe that the loss curve of the improved model on the training set and the verification set shows a better characteristic: the loss value decreases to the lowest and remains stable at the sixth epoch. In addition, the miou indicator has stabilized above 0.7 since the 20th epoch and has reached a maximum of 0.76. This result is significantly better than the traditional Unet model, showing that the improved model has stronger capability in feature extraction and edge recognition.

Finally, we apply two different versions of the Unet model to the test set for brain tumor image segmentation prediction. Through comparative analysis, it can be clearly seen that the improved Unet algorithm based on coordinate attention mechanism and ASPP module is obviously better than the traditional Unet in segmentation effect. This is not only reflected in the overall segmentation accuracy, but also in the edge detail processing shows a better ability. This result fully proves that our proposed method can effectively improve the performance of brain tumor image segmentation task, and provides a new idea for subsequent related research.

To sum up, this study effectively improves the traditional Unet by introducing coordinate attention mechanism and ASPP module, and makes it show more powerful performance in brain tumor image segmentation. This achievement not only provides a new technical means for the development of medical image analysis, but also provides a valuable reference for future related research. It is hoped that more advanced technologies will continue to be explored in the future to further promote the development of medical image processing.